\documentclass[final,5p,times,twocolumn,groupaddress]{elsarticle}

\usepackage{amssymb}
\usepackage{graphicx,amsmath}
\usepackage{natbib}
\usepackage{mathtools}
\usepackage[english]{babel}
\biboptions{numbers,sort&compress}

\usepackage[breaklinks,colorlinks=true,citecolor={blue}, linkcolor= {blue}, urlcolor ={blue}]{hyperref}

\journal{Solid State Communications}

\begin{document}

\begin{frontmatter}




\title{Tuning the lasing threshold of quantum well exciton-polaritons under a magnetic field in Faraday geometry: a theoretical study}


\author[l1,l2]{Le Tri Dat}
\address[l1]{Division of Applied Physics, Dong Nai Technology University, Bien Hoa City, Vietnam}
\address[l2]{Faculty of Engineering, Dong Nai Technology University, Bien Hoa City, Vietnam}
\ead{letridat@dntu.edu.vn} 

\author[l3,l4]{Nguyen Dung Chinh} 
\address[l3]{Institute of Fundamental and Applied Sciences, Duy Tan University, 06 Tran Nhat Duat St., Dist. 1, Ho Chi Minh City 70000, Vietnam} 
\address[l4]{Faculty of Environmental and Natural Sciences, Duy Tan University, 03 Quang Trung St., Hai Chau, Da Nang 50000, Vietnam}

\author[l5]{Vinh N.T. Pham}
\address[l5]{Department of Physics \& Postgraduate Studies Office, Ho Chi Minh City University of Education, Ho Chi Minh City, Vietnam}

\author[p1,p2]{Vo Quoc Phong} \ead{vqphong@hcmus.edu.vn}
\address[p1]{Department of Theoretical Physics, University of Science, Ho Chi Minh City 70000, Vietnam}
\address[p2]{Vietnam National University, Ho Chi Minh City 70000, Vietnam}

\author[l6,l7]{Nguyen Duy Vy\corref{cor1}}
\address[l6]{Laboratory of Applied Physics, Science and Technology Advanced Institute, Van Lang University, Ho Chi Minh City, Vietnam}
\address[l7]{Faculty of Technology, School of Technology, Van Lang University, Ho Chi Minh City, Vietnam}
\ead{nguyenduyvy@vlu.edu.vn} \cortext[cor1]{Corresponding author.}

\begin{abstract}
Polariton lasing represents a promising pathway toward the development of ultralow-threshold lasers that operate without requiring population inversion. The application of a magnetic field to a quantum well (QW) microcavity can significantly modify exciton-polariton properties, offering a powerful means to control their condensation dynamics. In this work, we theoretically investigate how a perpendicular magnetic field (Faraday configuration) influences the lasing threshold of QW exciton-polaritons. By incorporating magnetic-field-induced modifications to the exciton effective mass and Rabi splitting, we reveal that the relaxation kinetics—and consequently, the lasing threshold—are strongly affected. Under low-wavenumber pumping, increasing the magnetic field raises the threshold, while under high-wavenumber pumping, the threshold is reached at much lower pump intensities. Moreover, simultaneous increases in both pump energy and magnetic field significantly enhance relaxation efficiency, resulting in a substantially larger population of condensed polaritons. These findings provide valuable insights into the tunability of exciton-polariton condensation via external magnetic fields and offer guidance for the design of next-generation, low-threshold polariton lasers.
\end{abstract}


\end{frontmatter}

\section{Introduction}
Exciton-polaritons are hybrid quasiparticles arising from the strong coupling between quantum well (QW) excitons and cavity photons within a semiconductor microcavity. These particles exhibit a bosonic nature, enabling macroscopic quantum phenomena such as Bose–Einstein condensation (BEC) and polariton lasing at relatively high temperatures~\cite{08Kasprzak_detuning, Deng2010rmp, Carusotto2013RMP, Rajan2016, 22ALNaghmaish_magnetic}. Polariton lasers have attracted increasing attention due to their low lasing threshold and ultrafast dynamics, making them promising candidates for low-power optoelectronic devices~\cite{Azzini2011APL, Wertz2009APL, Schneider2013Nature}.

Reducing the polariton lasing threshold remains a central goal. Strategies include optimizing material systems~\cite{Guillet2011APL, 16LaiZnO}, engineering microcavity structures~\cite{Weihs2003Semi}, and applying external perturbations such as strain, electric, or magnetic fields. In particular, a perpendicular (Faraday geometry) magnetic field can significantly modify exciton properties, including binding energy, effective mass, oscillator strength, and exciton-photon coupling~\cite{15Pietka_polaB, 96Berger_polaRabi, 97Armitage_Hopfield, 08Solnyshkov_polaB}. Experimental studies have observed energy shifts, Zeeman splitting, and modifications of the polariton dispersion under magnetic fields, providing valuable knobs to control condensation conditions~\cite{Vladimirova2010PRB,17Rousset_exp}. 
Theoretical investigations into magneto-polariton systems have considered both phenomenological and microscopic models~\cite{12Moskalenko_magnetic_anal, 2015PRBStpnicki}. However, most previous works focused either on the exciton or polariton spectrum, or on steady-state photoluminescence, without a comprehensive treatment of the condensation kinetics in the presence of a magnetic field.
\begin{figure}[!htb] \centering
\includegraphics[width=.47\textwidth]{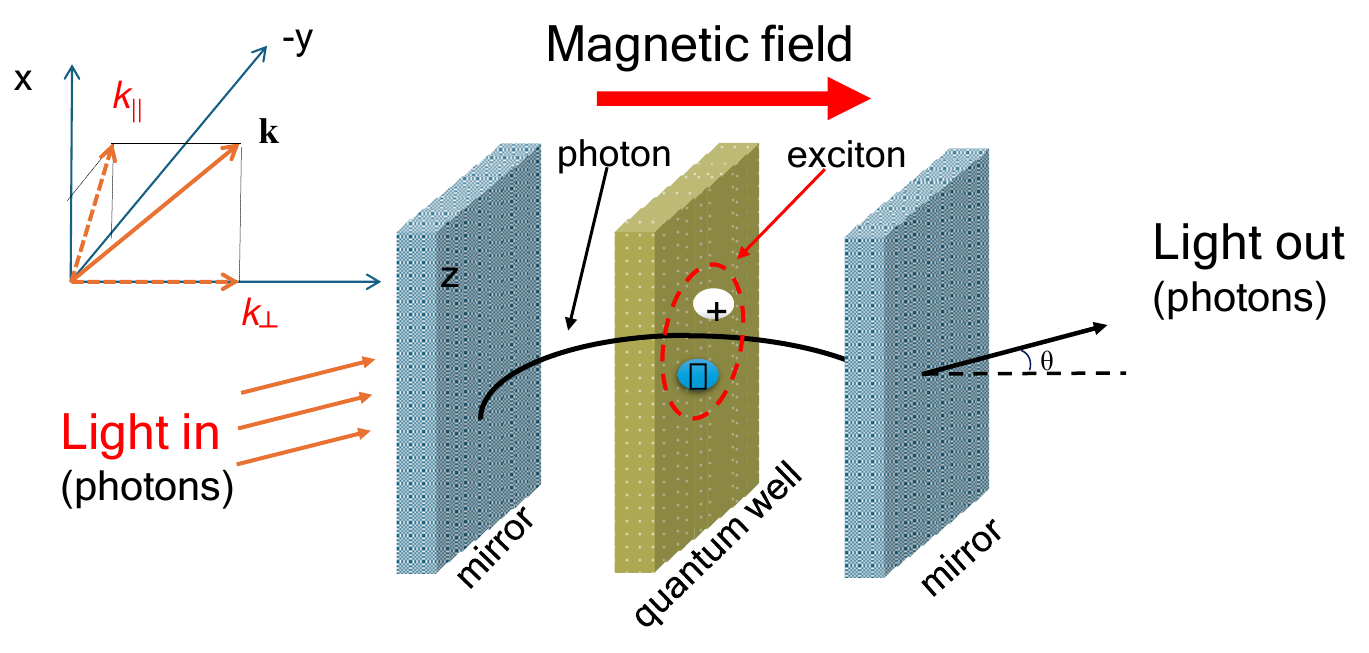} 
\caption{Model of the exciton-polariton in an optical microcavity. The magnetic field is perpendicular to the plane of the quantum well contaning the excitons..} \label{fig_model}
\end{figure}

In this work, we theoretically study how an external magnetic field influences the condensation threshold of QW exciton-polaritons, as shown in Fig. \ref{fig_model}, focusing on the interplay between exciton mass, Rabi splitting, and scattering-mediated relaxation. We consider a GaAs-based microcavity system and solve the Boltzmann kinetic equation to analyze how magnetic field strength and pump conditions affect polariton relaxation and condensation dynamics. Our results provide guidance for optimizing polariton lasing using magnetic-field tuning alone. 
The effects related to Zeeman splitting and spin \cite{97Armitage_Hopfield, 08Solnyshkov_polaB} are small, in the order of tens of $\mu$eV; therefore, will not be considered in this work.

Our theoretical model incorporates the influence of an external magnetic field on exciton energies and polariton dispersion, aiming to provide deeper insight into field-tunable polariton dynamics in quantum well systems. 
We begin by outlining the fundamental Hamiltonian framework used to describe polariton behavior, followed by a detailed analysis of how the lasing threshold depends on the magnetic field under different pump wavenumbers. 
The results of our theoretical investigation into the condensation process are then presented and discussed, culminating in a summary of the main conclusions.


\section{Exciton-polariton modified by magnetic field.}
The influence of magnetic fields on exciton properties has been extensively studied both experimentally and theoretically. 
Stepnicki et al. \cite{2015PRBStpnicki} showed that this shift can be expressed as ${\Delta{E}=D_2 B^2}$ where $D_2 \simeq$ 0.050 meV/T$^2$. 
However, to our knowledge, there are no published values of $D_2$ specifically for GaAs quantum wells from either theory or experiment.
In this work, we adopt system parameters from the GaAs/Ga$_{0.7}$Al$_{0.3}$As quantum wells studied by Bockelmann et al \cite{Bockelmann} and Bloch and Marzin \cite{97Bloch} assumed $D_2 = 0.085$ meV/T$^2$ to enhance the modeled influence of the magnetic field on polariton dispersion and kinetics. A similar value was used in the study by Stepnicki et al.~\cite{2015PRBStpnicki}, which investigated exciton-polaritons in a microcavity containing an 8 nm In$_{0.04}$Ga$_{0.96}$As QW placed between GaAs/AlAs distributed Bragg reflectors.

In this study, we adopted a higher value for $D_2$ , $D_2$ = 0.085 meV/$T^2$  to enhance the effect of the magnetic field on exciton-polariton dispersion and kinetics. 
%
For an increase in exciton mass, the relation ${\frac{1}{M(B)}=\frac{1}{M} -D_MB^2}$ where $D_M$ = (0.048 $\pm$ 0.002) $m_0^{-1}$T$^{-2}$ and $m_0$ is the mass of free electrons was also used. 

\begin{figure}[!htb] \centering
\includegraphics[width=.46\textwidth]{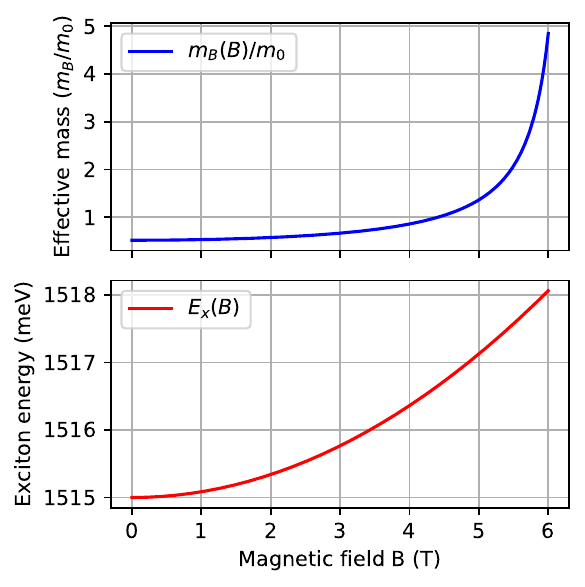} 
\caption{Change of exciton effective mass $m_B$ (top) and binding energy $E_x$ (bottom) versus the magnetic field.} \label{fig_mass-Ex}
\end{figure}

 The Rabi splitting increases with the magnetic field, $\Omega(B) = \Omega_0 (a_0/a(B))$ where $a_0$ and $a(B)$ are the Bohr radius without and with the magnetic field, respectively, was derived by Stepnicki et al. \cite{2015PRBStpnicki}, with a radius ratio,
${a(B) = {a_0}\sqrt{2}/\Big(
    1+\sqrt{1+3e^2a_0^4B^2/(2\hbar^2)}
    \Big)^{1/2}.
    }$
As a result of the exciton energy and mass modifications due to the magnetic field, as shown in Fig. \ref{fig_mass-Ex}, the exciton-polariton dispersion is significantly modified [see Eq. (\ref{pola})]. 

\begin{figure}[!htb] \centering
\includegraphics[width=.48\textwidth]{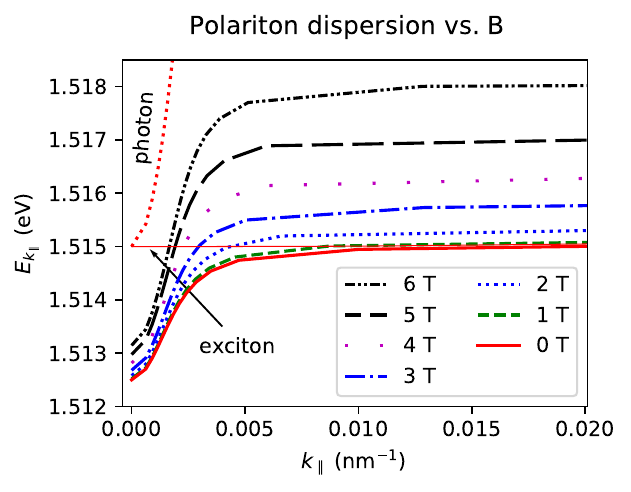} 
\caption{Polariton dispersion (thick red solid lines) resulting from the coupling between quantum well excitons (thin red solid lines) and cavity photons (red dotted lines). The detuning at $k_\parallel = 0$ is $\delta = E_x(0) - E_c(0) = 0$ meV. As the magnetic field increases, the polariton energy exhibits a nonlinear rise, with the effect being more pronounced in the plateau region ($k_\parallel > 0.005$ nm$^{-1}$) than at small $k$. At very high wavevectors, the dispersion becomes steeper, which significantly alters the relaxation dynamics.} \label{dispersion}
\end{figure}

Figure~\ref{dispersion} illustrates the variation of polariton energy as a function of the magnetic field $B$. As $B$ increases, the polariton energy also increases, with a notably stronger effect observed for $B > 2$ T. This energy shift is more pronounced in the high-$k$ region compared to the low-$k$ region ($k_{\parallel} \simeq 0$). As we will demonstrate in the following section, this modification in the polariton dispersion $E(k)$ significantly impacts both the scattering rates and the overall condensation dynamics.

To analyze the condensation kinetics, we begin with the Hamiltonian formalism describing the exciton, photon, and their mutual coupling. The total Hamiltonian is expressed as:
\begin{align}
H_{tot} =&H_{X} +H_{C} +H_{X-C},\label{Htotal}
\end{align}
where $B_k$/$B_k^\dagger$
 are the annihilation/creation operator of exciton, $b_k$/$b_k^\dagger$ are that of photon, and $H_{X-C}$ is the coupling between them (with a coupling strength $\Omega_X$)\cite{05PRBDoanTD,2009PRBVy,24SSCVy},
$H_X=
\hbar\omega_k^X
B_k^\dagger B_k$,  $H_C=\hbar\omega_k^C b_k^\dagger b_k$, $H_{X-C}= i\hbar\frac{\Omega_X}{2}(b_k^\dagger B_k -B_k^\dagger b_k)$.
The energy of the exciton (X) and photon (C) in a quantum well with a dielectric constant $\epsilon_b$ are 
\begin{align}
\hbar \omega^X_k = \hbar\omega_t +\frac{\hbar^2 k^2}{2M_x}, \mbox{ } \hbar\omega^C_k = \hbar\Big(\omega_0^2 +\frac{c^2 k^2}{\epsilon_b}\Big)^{1/2}, \label{ExEc}
\end{align}
The two-dimensional wave vector $k_\parallel$ has been written as $k_\parallel=k$ for brevity. From Eq. (\ref{Htotal}) a quasi-particle called the exciton-polariton could be obtained by diagonalizing the Hamiltonian, $H_{pol} =\sum\nolimits_k \omega_k^{pol} a_k^\dagger a_k$,
\begin{align}
2\omega_{k\pm}^{pol} =& \omega_k^X +\omega_k^C \pm \sqrt{(\omega_k^X-\omega_k^C)^2+\Omega_X^2}. \label{pola}
\end{align}
Here, the symbols $\pm$ denote the upper and lower polariton branches, respectively. Since the lower polariton branch, $\omega_k^-$, plays a dominant role in the relaxation dynamics, our analysis of the condensation kinetics focuses primarily on this dispersion. 

To examine the polariton kinetics, the polariton-phonon (p-ph) interaction ($H_{def}$) via a deformation potential and polariton-polariton (p-p) interaction ($H_{X-X}$) are taken into account, 
\begin{align}
    H_{tot} = H_{pol}  +H_{def} +H_{X-X}
\end{align}
where
\begin{align}
H_{def}=\sum_{q,q_z,k} [G(q,q_z) B_k^{\dagger}B_{k-q}
(c_{q,q_z}+c^{\dagger}_{-q,q_z}) +c.c], 
\end{align}
$c_\textbf{q}$ is the annihilation operator of phonons. The 3D wave vector of the phonons is split to the in-plane and the normal component with respect to the quantum well plane, $\textbf{q}$ = $(q,q_z)$ with $|\textbf{q}|=\sqrt{q^2+ q_z^2}$. The coupling strength $G$ is \cite{04PRBCaoHT,05PRBDoanTD,06PRBDoanTD}
,
\begin{align}
G(q,q_z)=& i\sqrt{\frac{\hbar |\textbf{q}|}{2V\varrho
u}}\frac{8\pi^2}{q_zL_z(4\pi^2-q_z^2L_z^2)}\sin(\frac{q_zL_z}{2}) 
\times \nonumber\\& \times 
\bigg\{a_e[1+\frac{b_e^2}{4}]^{-3/2}+a_h[1+\frac{b_h^2}{4}]^{-3/2}\bigg\}.\label{Gqqzp}
\end{align}
Here, $\rho$ is the material density, $u$ is the sound velocity, and $V = S L_z$ is the volume of the quantum well, with $S$ denoting the QW area and $L_z$ its thickness. The deformation potentials $a_e$ (for electrons) and $a_h$ (for holes) are taken from Ref.~\cite{2009PRBVy}.

The exciton-exciton interaction is modeled by $H_{X-X} = 6E_B a_B^2 x_{k+q} x_{k'-q} x_{k'} x_k$, where $E_B$ is the Coulomb exchange energy between two QW excitons, $a_B$ is the Bohr radius of the 2D exciton, and $x_k$ is the Hopfield coefficient that quantifies the excitonic weight of the polariton state. The polariton operator is given by $a_k = a_{k_\parallel} = x_{k_\parallel} B_{k_\parallel} + c_{k_\parallel} b_{k_\parallel}$, where the Hopfield coefficients satisfy: 
$\displaystyle c_{k_\parallel}^2={\frac{\omega_k^{pol}-\omega_k^X}{2\omega_k^{pol}-\omega_k^X-\omega_k^C}}$ and $\displaystyle x_{k_\parallel}^2={\frac{\omega_k^{pol}-\omega_k^C}{2\omega_k^{pol}-\omega_k^X-\omega_k^C}}$. These Hopfield coefficients play a crucial role in determining the relaxation dynamics, as they strongly influence the polariton-polariton scattering probabilities [see Eq.~(\ref{dn_dt}) below].






The relaxation kinetics of the exciton-polariton 
could be
described using the Boltzmann equations \cite{04PRBCaoHT, 2009PRBVy}. The time evolution of the polariton number $n_{\vec{k}}(t)$ is written as,
\begin{align}
\frac{\partial n_{\vec{k}}}{\partial t} =& p({\vec{k}},t) 
+\frac{\partial n_{\vec{k}}}{\partial t}|_{p-p} 
+
\frac{\partial{n_{\vec{k}}}}{\partial{t}}|_{p-ph}
- \frac{n_{\vec{k}}}{\tau_{\vec{k}}}     , \\
\frac{\partial n_0}{\partial t} =& \frac{\partial n_0}{\partial t}|_{p-p} 
+ \frac{\partial{n_0}}{\partial{t}}|_{p-ph}
- \frac{n_0}{\tau_0}     , 
\end{align}
where the quasi-stationary pump term $p({\vec{k}},t)$ play the role of external optical pumping to create and maintain the polariton in the system. $\tau_{\Vec{k}}$ is the polariton life-time which is weighted based on the life time of the exciton and the photon \cite{97Bloch}. 
 1/$\tau_{\Vec{k}}$ is used as Bloch and Marzin for a GaAs
quantum well: 1/$\tau_{\Vec{k}}$ = $c_{k}^2/\tau_c$ for 0 $<k<k_{cav}$ = 6$\times 10^{-4}$ cm$^{-1}$. At $k_{cav}$ the losses are determined by the
excitons coupled to a radiation continuum: 1/$\tau_{\Vec{k}}=1/\tau_x$ for $k_{cav}< k< k_{rad} = n_{cav}E_0^x/(\hbar c) =2.3\times 10^5$ cm$^{-1}$. $\tau_x$ and $\tau_c$ are the lifetimes of excitons and photons, respectively. 

The polariton-polariton scattering has the form, 
\begin{align}
\frac{\partial n_{\vec{k}}}{\partial t}|_{p-p}
=
& -\sum\limits_{\vec{k}, \vec{k}_1, \vec{k}_2} w^{p-p}_{\vec{k}, \vec{k}', \vec{k}_1, \vec{k}_2} [n_{\vec{k}} n_{\vec{k}'} (1+n_{\vec{k_1}})(1+n_{\vec{k_2}}) 
- n_{\vec{k}_1} n_{\vec{k}_2} (1+n_{\vec{k}})(1+n_{\vec{k}'})],  
\end{align}
where $\vec{k}=\vec{k_1}-\vec{q}$ and $\vec{k'}=\vec{k_2}+\vec{q}$. The probability of p-p transition is \cite{99Tassone},

\begin{align}
\frac{\partial n_{\vec{k}}}{\partial t} = 
\frac{\pi}{\hbar}
\frac{S^2}{(2\pi)^4}
\frac{\Delta E^2|M|^2 u_k^2 u_{k'}^2 u_{k_1}^2 u_{k_2}^2}{
\frac{\partial^2 E(k')}{\partial k'^2}
\frac{\partial^2 E(k_1)}{\partial k_1^2}
\frac{\partial^2 E(k_2)}{\partial k_2^2}
}
R(k,k',k_1,k_2), \label{dn_dt}
\end{align}
where 
\begin{align}
& R(k,k',k_1,k_2) =\int dq^2 \Big\{ 
\frac{1}{\sqrt{
[(k+k_1)^2-q^2] [q^2-(k-k_1)^2]}}
\times \nonumber \\& \times
\frac{1}{\sqrt{[(k'+k_2)^2-q^2][q^2-(k'-k_2)^2]}}
\Big\}
\end{align}
and 
\begin{align}
M\simeq 2\sum_{k,k'} V_{\vec{k}-\vec{k'}} (\phi_k^2-\phi_k\phi_{k'})
\simeq6E_0(B) a_0(B)^2/S,
\end{align}
where $a_0$ is the exciton Bohr radius, $E_0$ the binding energy, and $\phi_k$ the wave function.
Therefore, a change in the polariton energy $\omega_k(B)$ due to the magnetic field B gives rise to the change in the polariton-polariton scattering rate via the changes in $\Delta E$ and $M$, and so does the number $\partial^2 E/\partial^2 k^2$ and $R(k ...)$.  
The polariton-phonon (p-ph) scattering term is
\begin{align}
\frac{\partial n_{\vec{k}}}{\partial t}|_{p-ph}
=&
-\sum\limits_{\vec{q}, \sigma=\pm 1} w^{p-ph}_{\vec{k}, \vec{q}, \sigma} [n_{\vec{k}} (1+n_{\vec{k}+\vec{q}}) N_{q,\sigma} 
- n_{\vec{k}+\vec{q}} (1+n_{\vec{k}}) N_{q,-\sigma}],  
\end{align}
where $N_{q,\sigma}=N_q+1/2+\sigma/2$ and $N_q$ is the Bose distribution of phonons.  The polariton-phonon transition probability is \begin{align}
w_{\vec{k},\vec{k'}} =
&\frac{L_z(u_ku{k'}\Delta_{k,k'})^2}{\hbar\rho Vu^2q_z}
B^2(q_z) D^2(|\vec{k}-\vec{k'}|)
N^{ph}_{E_{k}-E_{k'}} \theta (\Delta_{\vec{k},\vec{k'}}-|\vec{k}-\vec{k'}|),
\end{align}
where
\begin{align}
\Delta_{\vec{k},\vec{k'}}=& |E_{k'}-E_k|/(\hbar u),  
q_z =\sqrt{\Delta_{\vec{k},\vec{k'}}^2-|\vec{k}-\vec{k'}|^2}, 
\\
D(q)=& d_eF(\frac{qm_h}{m_e+m_h})-d_hF(\frac{qm_e}{m_e+m_h}),
\\
B(q) =& \frac{8\pi^2}{L_zq(4\pi^2-L_z^2q^2)} \sin(L_zq/2),
F(q) = (1+(qa_0/2)^2)^{1/3}.
\end{align}

In this study, a quasi-stationary pump is assumed 
\begin{align}
p_c(\Vec{k}, t) = p_0 e^{-\frac{1}{2} \big[\frac{E(k) -E(k_p)}{\Gamma} \big]^2}\tanh(t/t_0),
\end{align}
where $p_0$ is the pump power, $\Gamma$ = 0.25 meV is the energy width, $t_0$ = 50 ps, and $k_p$ is the wavenumber {at the center of the pumped exciton reservoir, which is large enough to avoid coherent parametric scattering into the lowest state \cite{05PRBDoanTD}}. 

\section{Results and discussion}
Here, to simulate the dynamics, we used the following parameters for a GaAs quantum well: area of the QW, $S$ = 100 $\mu$m$^2$; thickness, $L_z$ = 5 nm; background dielectric constant, $\epsilon_b$ = 11.9; effective masses of electrons and holes~\cite{Tan2019PLA, Vy2020Superlattices, 22ALNaghmaish_magnetic}: $m_e = 0.067m_0$, $m_h = 0.45m_0$, where $m_0$ is the free electron mass; Rabi splitting, $\Omega_X = 5$ meV; and cavity photon energy, $\hbar\omega_0 = 1.515$ eV. The photon and exciton lifetimes are taken from Ref.~\cite{97Bloch}, with $\tau_c = 4$ ps and $\tau_x = 20$ ps. The system temperature is fixed at 4 K.

\begin{figure*}[!ht] \centering
\includegraphics[width=.9\textwidth]{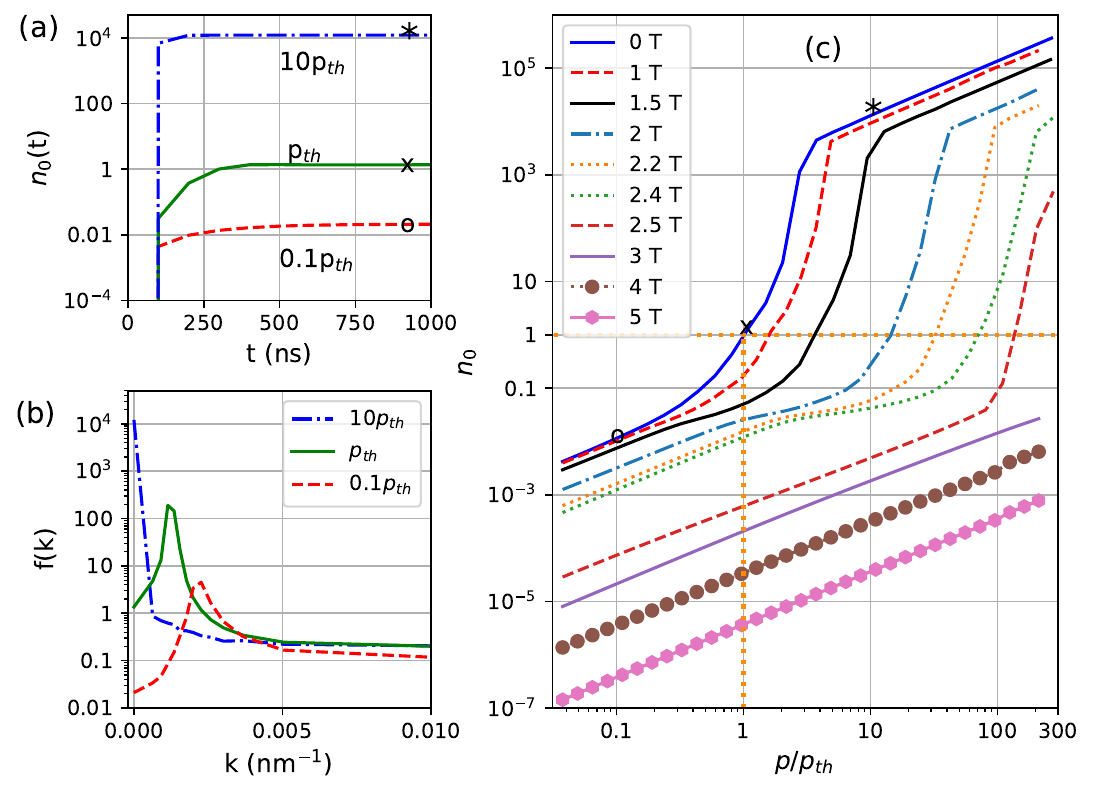} 
\caption{((a) Time evolution of the condensed polariton population ($n_0$) for three representative pump strengths at $B = 0$. (b) Corresponding momentum distribution function $f(k)$ at the saturated states, where $n_0(t)$ becomes constant. The three cases shown in (a) and (b) are marked by ‘o’, ‘x’, and ‘*’ symbols in (c), which plots the S-shaped dependence of $n_0$ on the pump strength $p/p_{\text{th}}$. Here, $k_p = 0.02$ nm$^{-1}$. At this low-$k$ pumping regime, increasing the magnetic field $B$ leads to inefficient condensation. For $B > 2$ T, much higher pump strengths are required to reach the condensation threshold ($n_0 > 1$), as indicated by the horizontal orange dotted line. For $B > 3$ T (purple solid line), condensation is no longer achievable, even at very high pump intensities.} \label{nt_B}
\end{figure*}

The relaxation process gradually builds up the number of condensed polaritons, denoted by $n_0$. We define the pump threshold, $p_{\text{th}}$, as the pump strength that yields $n_0 \simeq 1$ in the steady state. All other pump strengths are then expressed in units of $p_{\text{th}}$. First, we consider a pump in the low-wavenumber region, $k_p = 0.02$ nm$^{-1}$. Below threshold, for instance at $p = 0.1p_{\text{th}}$, the number of condensed polaritons remains very low, $n_0 \simeq 0.02$, at the stationary state after 1 ps, as shown by the red dashed line with ‘o’ markers in Fig. \ref{nt_B}(b). 
At threshold, where $n_0 \simeq 1$ (green solid line), the distribution still shows a significant population at $k > 0$, indicating that a large number of polaritons accumulate in this region—a phenomenon commonly referred to as the bottleneck effect. It has been demonstrated that enhanced polariton–polariton (p–p) scattering can effectively suppress this effect, enabling a clearer condensation profile that aligns well with the Bose–Einstein distribution~\cite{05PRBDoanTD,06nature}.

Far above the threshold, at $p = 10p_{\text{th}}$, the condensate number reaches $n_0 \sim 10^4$ (blue dash-dotted line marked with ‘*’), signifying a well-formed Bose–Einstein condensate. These three representative cases of $n_0$ versus $p$, along with intermediate pump values, form an S-shaped curve as shown in Fig.~\ref{nt_B}(c) (blue solid line). %
In a recent experiment, Rousset et al.~\cite{17Rousset_exp} demonstrated that by reducing the photon–exciton detuning to $\delta = E_x(0) - E_c(0) = -8.5$ meV, condensation could still be achieved under a relatively low magnetic field of 2 T. In our present study, to isolate and clarify the role of the magnetic field, we fix the detuning at $\delta = 0$, and systematically investigate the condensation behavior solely as a function of the magnetic field strength.
 
Increasing the magnetic field under low-$k$ pumping conditions leads to inefficient condensation. In Ref.~\cite{Wertz2009APL}, the magnetic field was shown to split the orthogonal pseudospin states of excitons in the quantum well, thereby shifting their energies. As a consequence, one of the pseudospin components becomes off-resonant with the cavity photon mode, effectively reducing the density of exciton states that can couple to light. This results in a lowering of the condensation threshold. However, in the present study, the Boltzmann equation used does not account for (pseudo)spin degrees of freedom.

\begin{figure*}[!ht] \centering
\includegraphics[width=1.02\textwidth]{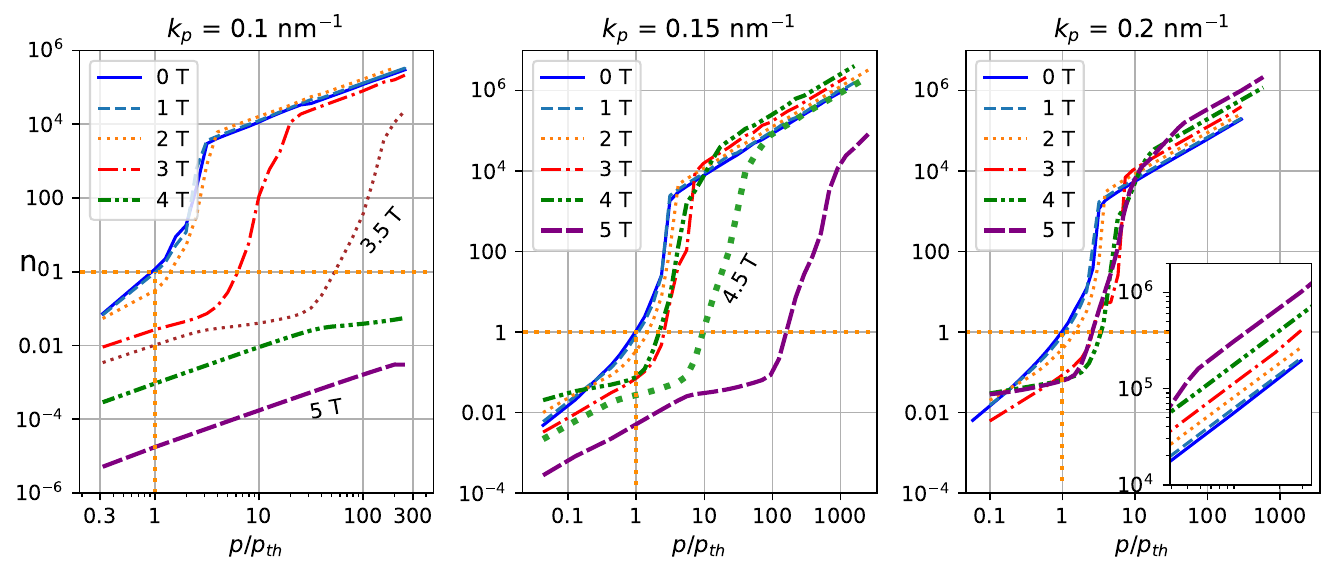} 
\caption{S-shaped curves showing the dependence of the condensed polariton number $n_0$ on the normalized pump strength $p/p_{\text{th}}$ for various magnetic fields and pump wavenumbers. For $k_p = 0.1$ nm$^{-1}$ (left panel), condensation remains inefficient when $B > 3.5$ T. Increasing the pump energy by raising $k_p$ enhances the condensation process. At $k_p = 0.15$ nm$^{-1}$ (middle panel), a moderate magnetic field of $B = 3$ T (green dash-dot-dotted line) enables effective condensation with a threshold near $p \approx 3p_{\text{th}}$. For even higher pumping at $k_p = 0.2$ nm$^{-1}$ (right panel), strong condensation is achieved, with the condensed population reaching $n_0 > 10^6$ (purple dashed line).} \label{3subs}
\end{figure*}

The impact of pump energy is illustrated in Fig.~\ref{3subs}, where we consider $k_p = 0.1$, 0.15, and 0.2 nm$^{-1}$. Compared to the case of $k_p = 0.02$ nm$^{-1}$, the lasing threshold is substantially lower. At $B = 3$ T, the threshold is only about eight times higher than at $B = 0$. However, even a modest increase in field to $B = 3.5$ T leads to a significant reduction in condensation efficiency, and $n_0 > 1$ is achieved only at very high pump strengths—up to 50 times that of the zero-field case (brown dotted line).

At even higher pump energies, such as $k_p = 0.15$ nm$^{-1}$ (middle panel), condensation efficiency is further enhanced. The S-shaped curve for $B = 4$ T (green dash-dot-dotted line) lies close to those for lower field values ($B = 0$–3 T), and the peak value of the condensed population becomes noticeably higher than that at lower $k_p$.

At very high-$k$ pumping, the influence of the magnetic field on condensation becomes more pronounced, as it significantly enhances the condensed polariton number $n_0$. With a moderate pump strength—only 20 to 30 times the threshold—a high condensate density of $n_0 \simeq 10^6$ can be achieved at $B = 5$ T (violet dashed line). The inset further illustrates that the increase in $n_0$ outpaces the increase in magnetic field strength, as seen by the widening separation between the purple dashed line and the green dash-dot-dotted line compared to other field values. These results suggest that combining high-wavenumber pumping with a moderate magnetic field enables highly efficient condensation. This provides useful guidance for experimental implementation of high-power polariton lasing systems.

\section{Summary}
In summary, we have theoretically investigated the condensation kinetics of quantum well exciton-polaritons under the influence of a perpendicular magnetic field. While our model is applied specifically to a GaAs quantum well system, the calculation method is general and can be extended to other materials with similar structures and polariton dispersions.

Our results demonstrate that, under low-wavenumber pumping, the magnetic field hinders condensation, leading to a significant increase in the lasing threshold. In contrast, under high-wavenumber pumping, the magnetic field enhances the relaxation from high-$k$ to low-$k$ states, thereby promoting more efficient condensation. This behavior leads to a nonlinear increase in the condensed polariton population with increasing magnetic field strength.

In contrast to previous studies where efficient condensation required the combination of multiple factors—such as negative detuning and magnetic field—we show that magnetic tuning alone can effectively control the condensation threshold. These findings offer valuable insights into the role of magnetic fields in optimizing exciton-polariton lasing, and may serve as a guideline for designing low-threshold, high-efficiency polariton laser systems.



\bibliographystyle{elsarticle-num}
\bibliography{pola_Slike.bib}

\end{document}